\documentclass[11pt]{article}
\usepackage[utf8]{inputenc}
\usepackage{mathtools,amsmath,amssymb}
\usepackage{graphicx}
\usepackage{lmodern}
\usepackage[T1]{fontenc}
\usepackage{microtype}
\usepackage[pagebackref=false]{hyperref}
\numberwithin{equation}{section}
\usepackage{dsfont}

\usepackage[textsize=tiny]{todonotes}

\usepackage{xspace}
\usepackage{bibentry}
\usepackage{easybmat}

\usepackage{graphicx}

\usepackage{subcaption} 

\usepackage[numbers,sort&compress]{natbib}
\usepackage{hypernat}
 
\usepackage{tikz}
\usetikzlibrary{decorations.pathreplacing} 
 
\usepackage[margin=2.2cm]{geometry} 
    
\newcommand{\de}{\mathrm{d}}
\newcommand{\gam}{\gamma}
\newcommand{\br}[1]{[#1]_\gam}

\newmuskip\pFqmuskip

\newcommand*\pFq[6][8]{%
  \begingroup 
  \pFqmuskip=#1mu\relax
  \mathchardef\normalcomma=\mathcode`,
  \mathcode`\,=\string"8000
  \begingroup\lccode`\~=`\,
  \lowercase{\endgroup\let~}\pFqcomma
  {}_{#2}\phi_{#3}{\left[\genfrac..{0pt}{}{#4}{#5};#6\right]}%
  \endgroup
}
\newcommand{\pFqcomma}{{\normalcomma}\mskip\pFqmuskip}

\newcommand{\vm}{\vec{m}}

\begin{document} 

\begingroup
\begin{center}

\vspace{5em}
 \begingroup\LARGE
\bf The steady state of the boundary-driven multiparticle asymmetric diffusion model
\par\endgroup
 \vspace{3.5em}
 \begingroup\large \bf
 {Rouven Frassek$^{\,a}$  and Istv\'{a}n M. Sz\'{e}cs\'{e}nyi$^{\,a,\,b}$}
 \par\endgroup
\vspace{2em}

\begin{center} 
$^{\,a}$ University of Modena and Reggio Emilia, FIM,\\
Via G. Campi 213/b, 41125 Modena, Italy
\\[0.1in]
$^{\,b}$ Nordita,  KTH Royal Institute of Technology and Stockholm University,\\  Hannes Alfvéns väg 12,  SE-106  91  Stockholm,  Sweden
\\[0.6in]
 \end{center}

\vspace{-2em}

\end{center} 
\begin{abstract} 
We consider the multiparticle asymmetric diffusion model (MADM) introduced by Sasamoto and Wadati with integrability preserving reservoirs at the boundaries. In contrast to the open asymmetric simple exclusion process (ASEP) the number of particles allowed per site is unbounded in the MADM.  
Taking inspiration from the stationary measure in the symmetric case, i.e. the rational limit, we first obtain the length $1$ solution and then show that the steady state can be expressed as an iterated product of Jackson q-integrals. 
In the proof of the stationarity condition, we observe a cancellation mechanism that closely resembles the one of the matrix product ansatz.
To our knowledge, the occupation probabilities in the steady state of the boundary-driven MADM were not available before.
\end{abstract}


\addtocontents{toc}{\protect\setcounter{tocdepth}{2}}

\thispagestyle{empty}

\section{Introduction}
A quarter century ago, the multiparticle asymmetric diffusion model (MADM) was introduced by Sasamoto and Wadati \cite{Sasamoto} as an integrable generalisation of the celebrated asymmetric simple exclusion process (ASEP).  Similar to the ASEP, the model is defined on a one-dimensional lattice with periodic boundary conditions, and particles jump at a certain rate to nearest-neighbouring sites. The asymmetry in the rates of left and right jumps is governed by a parameter $0<\gamma<1$. The occupation number of particles per site is unbounded in the  MADM and multiple simultaneous jumps are allowed. The jump rates only depend on the number of particles that jump to the left or to the right. As such the model belongs to the class of zero-range processes with factorised steady state, c.f.~\cite{2004JPhA...37L.275E,2006JPhA...39.1565G,Povolotsky,2015arXiv150700778F}. 

The MADM  is solvable by Bethe ansatz, which can be explained by the fact that it can be mapped to the XXZ spin chain with non-compact infinite-dimensional spin representations \cite{Frassek:2019isa}. In particular, the q-deformation parameter of the underlying $U_q(sl_2)$ algebra is identified as the asymmetry parameter via $q^2=\gam$. This mapping parallels the relation between the ASEP and the ordinary XXZ Heisenberg spin chain. More precisely, the Markov generator of the process is mapped to an integrable Hamiltonian that, within the quantum inverse scattering method, is part of the transfer matrix; see \cite{Crampe:2014aoa} for an overview. One advantage of such formulation is that for a given integrable particle process, boundary reservoirs can be introduced without breaking the integrable structure by following the work of Sklyanin \cite{Sklyanin:1988yz}. This strategy led to the definition of the MADM with integrable boundary reservoirs  \cite{Frassek:2022fjs}. 

In the rational limit of $\gamma\to 1$, the MADM with integrable boundary reservoirs (boundary driven MADM) reduces to the open harmonic process defined in \cite{Frassek:2019vjt} with non-compact spin value $s=1/2$ for where particles jump symmetrically to the left and right neighbouring sites. Remarkably, this model has an absorbing dual process and, therefore, can be mapped to equilibrium by a non-local transformation, which in particular allows us to compute the steady state exactly \cite{Frassek:2021yxb}. Using the integral representation of the Beta function, this closed-form expression for the steady state was written in the form of a nested integral in \cite[Appendix A]{2023arXiv230714975C}, see also \cite{2023arXiv230714975C,2023arXiv230702793C} where it has been interpreted as a mixed measure. We further remark that a  similar approach to the symmetric simple exclusion process (SSEP) has been presented in \cite{Frassek:2019imp,Frassek:2020omo}. 
However, unfortunately, the pathway outlined above has not yet been fully developed for the asymmetric case (XXZ-type models). For some recent interesting developments in this direction for the ASEP, we refer the reader to \cite{2023JPhA...56A4001S,2022arXiv221207349B}.

In this paper, we provide a shortcut to the stationary measure of the boundary-driven MADM by taking inspiration from the results available in the rational limit. We introduce the boundary-driven MADM in Section~\ref{sec:qhahn}   along with the stationarity condition. Section~\ref{sec:ansatz} contains the stationary measure for length $N=1$ (see Appendix~\ref{app:l1} for the proof)  and presents the conjecture for general $N$, see \eqref{eq:stead} for our main result. The conjecture is based on the observation that the length $1$ steady state can be written as a Jackson q-integral.  
In Section~\ref{sec:steady}, we prove the conjecture and reveal a term-by-term cancellation as common for the matrix product ansatz.
In the case of equilibrium, where the insertion and extraction rates at both boundaries coincide, we recover a product measure as outlined in Section~\ref{sec:equi}. Finally, we end with some concluding remarks in Section~\ref{sec:con}. Some useful formulas regarding the q-calculus and a generating function method are collected in Appendix~\ref{app:qcalc} and \ref{app:genf}, respectively.

\section{The boundary-driven MADM}\label{sec:qhahn}  
The boundary-driven multiparticle asymmetric diffusion model (MADM) is a continuous-time Markov process of interacting particles defined on a one-dimensional lattice of length $N$. The number of particles per lattice site is unbounded such that the occupation number $m_i\in \mathbb{N}_0$ at site $i=1,\ldots,N$. The particles can jump from a given site to the two neighbouring sites with rates that depend on the number of particles jumping and the asymmetry parameter $0<\gam<1$. In Section~\ref{sec:process}, we present the process that is obtained from the stochastic integrable Hamiltonian \cite{Frassek:2022fjs} in Section~\ref{sec:stoch}. The stationarity condition is found in Section~\ref{sec:statc}.

\subsection{The process}\label{sec:process}
The MADM is defined through the action of the  Markov generator $\mathcal L$ on functions $f(\vm)$ with $\vm=(m_1,\ldots,m_N)$ as follows
\begin{equation}
\label{eq:gen}
{\mathcal L}f(\vm) := {\mathcal L}_L f(\vm)  + \sum_{i=1}^{N-1}{\mathcal L}_{i,i+1}f(\vm)  + {\mathcal L}_R f(\vm) 
\end{equation}
where the density of the Markov generator acts as
\begin{equation}
\label{eq:genbulk}
{\mathcal L}
_{i,i+1} f(\vm) 
=  
\sum_{k=1}^{m_i}\frac{1}{\br{k}} \Big[f(\vm-k \delta_i + k \delta_{i+1}) - f(\vm)\Big] +
\sum_{k=1}^{m_{i+1}} \frac{\gam^k}{\br{k}} \Big[f(\vm+k\delta_i - k \delta_{i+1}) - f(\vm)\Big]\,,
\end{equation}
and, for $i\in\{1,N\}$, at the boundaries we have
\begin{equation}
\label{eq:genbdryL}
{\mathcal L}
_{L} f(\vm) =
\sum_{k=1}^{m_1} \frac{\gam^k}{\br{k}}\Big[f(\vm-k\delta_1) - f(\vm)\Big]
 +   
\sum_{k=1}^{\infty}\frac{ \beta_L^k}{\br{k}} \Big[f(\vm+k\delta_1) - f(\vm)\Big]\,,
\end{equation}
and
\begin{equation}
\label{eq:genbdryR}
{\mathcal L}
_{R} f(\vm) =
\sum_{k=1}^{m_N} \frac{1}{\br{k}}\Big[f(\vm-k\delta_N) - f(\vm)\Big]
 +   
\sum_{k=1}^{\infty} \frac{(\gam\beta_R)^k}{\br{k}} \Big[f(\vm+k\delta_N) - f(\vm)\Big]\,.
\end{equation} 
Here, we denote by $\delta_i$ the elementary unit vector 
\begin{equation}
\delta_i=(\underbrace{0,\ldots,0,1}_{i},\underbrace{0,\ldots,0}_{N-i})\,,
   \end{equation}   
and introduced the q-number as
\begin{equation}\label{eq:qnum}
    \br{k}=\frac{1-\gam^k}{1-\gam}\,.
\end{equation} 
The two boundary parameters take values $0<\beta_L,\beta_R<1$.

We like to stress that integrability fixes the boundary terms only up to a rescaling in the boundary parameters. The choice in \eqref{eq:genbdryL} and \eqref{eq:genbdryR} is made such that the system is in equilibrium for $\beta_L=\beta_R$ which is discussed in further detail in Section~\ref{sec:equi}.

\subsection{Stochastic Hamiltonian}\label{sec:stoch}

The generator of the process \eqref{eq:gen} is obtained from the stochastic Hamiltonian given below using the relation ${\mathcal L}f(\vm)=-\sum_{\vm'}f(\vm') \langle \vm'|H|\vm\rangle$, see e.g.~\cite{2015JSP...161..821B}. In the vectorial notation, the $f(\vm)$ function is represented  as $f(\vm)=\langle f|\vm\rangle$ with $\langle f|=\sum_{\vm'}f(\vm') \langle \vm'|$.  The stochastic Hamiltonian is of nearest-neighbour type and can be written as
\begin{equation}\label{eq:ham}
    H=B_L+\sum_{i=1}^{N-1} \mathcal{H}_{i,i+1}+B_R\,.
\end{equation}
Here $\mathcal{H}_{i,i+1}$ is the Hamiltonian density acting non-trivial on sites $i$ and $i+1$ and the boundary terms $B_L$ and $B_R$ that act non-trivially on the first and last site respectively.

The Hamiltonian acts on the $N$-fold tensor product of infinite-dimensional highest weight modules denoted by $|m\rangle $ with $m=0,1,2,\ldots$ such that the configuration space is described by orthogonal infinite-dimensional vectors $|\vm\rangle=|m_1\rangle \otimes\ldots\otimes |m_n\rangle$. 
The action of the  Hamiltonian density on two neighbouring sites is of the form
\begin{equation}\label{eq:hact}
\begin{split}
\mathcal{H}|m\rangle \otimes |m'\rangle=\left(\sum _{k=1}^{m} \frac{1}{\br{k}}+\sum _{k=1}^{m'} \frac{\gamma^{k}}{\br{k}}\right)|m\rangle \otimes |m'\rangle&-\sum_{k=1}^m\frac{1}{\br{k}}|m-k\rangle \otimes |m'+k\rangle\\&-\sum_{k=1}^{m'} 
\frac{\gamma ^{k} }{\br{k}} |m+k\rangle \otimes |m'-k\rangle\,.
 \end{split}
\end{equation}  
The boundary terms act non-trivially on the first and last site, respectively. Their action reads
\begin{equation}\label{eq:bd1}
\begin{split}
B_L|m\rangle =\left(\sum _{k=1}^{m} \frac{\gamma^{k}}{\br{k}}+\sum_{k=1}^\infty\frac{\beta_L^k}{\br{k}}\right)|m\rangle&
-\sum_{k=1}^m 
\frac{\gamma ^{k} }{\br{k}} |m-k\rangle -\sum_{k=1}^\infty\frac{\beta_L^k}{\br{k}}|m+k\rangle \,,
\end{split}
\end{equation} 
and
\begin{equation}\label{eq:bd2}
\begin{split}
B_R|m\rangle =\left(\sum _{k=1}^{m} \frac{1}{\br{k}} +\sum_{k=1}^\infty\frac{(\gam \beta_R)^k}{\br{k}}\right)|m\rangle&
-\sum_{k=1}^m\frac{1}{\br{k}}|m-k\rangle -\sum_{k=1}^\infty\frac{(\gam\beta_R)^k}{\br{k}}|m+k\rangle \,.
\end{split}
\end{equation} 
This formulation is equivalent to the one given in Section~\ref{sec:process}.

\subsection{Stationarity condition}\label{sec:statc}
We are interested in the stationary probability  measure $\mu(\vec m)$. The evolution of the probability measure $P(\vm)$ of the Markov process is described by the action of the transposed Markov generator $\mathcal{L}^t P(\vm)$ defined via $\mathcal{L}^t P(\vm)=\sum'_{\vm'}P(\vm')\mathcal{L}\,{\bf1}_{\vm}(\vm')$ with ${\bf1}_{\vm}(\vm')=\delta_{\vm,\vm'}$ and $\sum'_{\vm'}$ denoting the sum over all configurations $\vm'\neq\vm$. The stationarity of the probability measure then implies that $\mathcal{L}^t \mu(\vm)=0$, c.f.~\cite{2015JSP...161..821B}. 
In terms of the stochastic Hamiltonian the stationarity condition reads
\begin{equation}\label{eq:Hmu}
     H|\mu\rangle=0\,,
\end{equation}
where  $\mu(\vm)=\langle \vm|\mu\rangle$.
Explicitly this yields
\begin{equation}\label{eq:stateq}
\begin{split}
\left(\sum_{i=1}^N\sum_{k=1}^{m_i}\frac{1+\gam^k}{\br{k}}  +\sum_{k=1}^\infty\frac{(\gam\beta_R)^k+\beta_L^k}{\br{k}}\right)\mu(\vm)&=\sum_{k=1}^\infty \frac{\gam^k}{\br{k}} \mu(\vm+k\delta_1)+\sum_{k=1}^\infty \frac{1}{\br{k}} \mu(\vm+k\delta_N)\\&\quad
+\sum_{k=1}^{m_1}\frac{\beta_L^k}{\br{k}}\mu(\vm-k\delta_1)+\sum_{k=1}^{m_{N}}\frac{(\gam\beta_R)^k}{\br{k}}\mu(\vm-k\delta_{N})\\
&\quad+\sum_{j=1}^{N-1}
\sum_{k=1}^{m_j} \frac{\gam^k}{\br{k}} \mu(\vm-k\delta_{j}+k\delta_{j+1})\\
&\quad+\sum_{j=1}^{N-1}\sum_{k=1}^{m_{j+1}} \frac{1}{\br{k}} \mu(\vm+k\delta_j-k\delta_{j+1})\,.
\end{split}
\end{equation}  
In the following section, we will first determine the stationary measure $\mu(\vm)$ for length $N=1$ and present a conjecture for arbitrary length $N$. The proof of which is postponed to Section~\ref{sec:steady}.

\section{Stationary measure: from one site to all $N$}\label{sec:ansatz}
  
In order to construct the steady state, let us recall the result for the  rational $\gamma\to 1$ model, see \cite{Frassek:2021yxb}. For length $N=1$, the stationarity measure found simply reads
\begin{equation}
\lim_{\gam\to 1} \mu (m_1)=\frac{(\beta_L-1)(\beta_R-1)}{\beta_L-\beta_R}\sum_{k=m_1+1}^\infty \frac{\beta_R^k-\beta_L^k}{k} \,.
\end{equation} 
The prefactor is just a normalisation to ensure that $\sum_{m_1=0}^\infty \lim_{\gam\to 1} \mu(m_1)=1$. Further, we observe no mixing terms between $\beta_L$ and $\beta_R$. Assuming that no such mixing appears in the q-deformed case allows us to write an ansatz for the length $N=1$ solution. Inserting this ansatz into the stationarity condition \eqref{eq:stateq}, we obtain
\begin{equation}\label{eq:leng1}
 \mu (m_1)=\frac{1}{c_1}\sum_{k=m_1+1}^\infty \frac{(\gam\beta_R)^k-\beta_L^k}{\br{k}} \,,
\end{equation} 
with the q-numbers defined in \eqref{eq:qnum} and a normalisation constant $c_1=\sum_{m_1=0}^\infty \mu (m_1)$, see also Section~\ref{sec:steady}. The proof that \eqref{eq:leng1} obeys the stationarity condition can be found in Appendix~\ref{app:l1}. 

We will now rewrite the stationary measure for $N=1$ as given in  \eqref{eq:leng1} as a Jackson q-integral whose definition can e.g. be found in  \cite[(1.11.1)-(1.11.3)]{gasper2004basic}. Consider a function $g(t)$, the Jackson q-integral with boundaries $a,b$ is then defined via
\begin{equation}\label{eq:jack1}
\int_a^bg(t)\de_\gam t=\int_0^bg(t)\de_\gam t-\int_0^ag(t)\de_\gam t\,,
\end{equation} 
with 
\begin{equation}\label{eq:jack2}
 \int_0^ag(t)\de_\gam t=a(1-\gam)\sum_{n=0}^\infty g(a\gam^n)\gam^n\,.
\end{equation}  
Next, we notice that each term in the sum of \eqref{eq:leng1} can be written with some simple manipulations as follows: 
\begin{equation}\label{eq:rewritesuml0}
\begin{split}
 \sum_{k=n+1}^\infty \frac{\beta^k}{\br{k}}&=(1-\gam)  \sum_{\ell=0}^\infty \sum_{k=n+1}^\infty \beta^k\gam^{k \ell} = (1-\gam) \sum_{\ell=0}^\infty \sum_{k=0}^\infty \beta^{k+n+1}\gam^{(k+n+1) \ell} 
 = \beta(1-\gam)\sum_{\ell=0}^\infty  \frac{\left(\beta\gam^{ \ell}\right)^n}{1-\beta\gam^{\ell} }\gam^{  \ell}\,.
 \end{split}
\end{equation} 
Comparing this result with the definition of the Jackson integral then yields
\begin{equation}\label{eq:l1res}
     \mu (m)=\frac{1}{c_1}\int_{\beta_L}^{\gam\beta_R} \frac{t^{m} }{1- t}\de_\gam t\,.
\end{equation}
Remarkably,  this expression now resembles the expression for the rational case that was found in \cite{2023arXiv230702793C,2023arXiv230714975C}, up to a change of variables\,\footnote{In loc. cit. the density variables $\beta=\frac{\rho}{1+\rho}$ and $t=\frac{\theta}{1+\theta}$ are used. }! However, the standard integral is replaced by the Jackson q-integral, and $\gam$ is introduced in the upper integration limit.

This result, combined with our preliminary numerics at length $2$, suggests proceeding in analogy to the rational case and leads us to the following representation of the stationary measure of the MADM in terms of nested Jackson integrals:
\begin{equation}\label{eq:stead}
\begin{split}
 \mu (\vm) &=\frac{1}{c_N}\int_{\beta_L}^{\gam\beta_R}\de_\gam t_1\int_{t_1}^{\gam\beta_R}\de_\gam t_2\cdots \int_{t_{N-1}}^{\gam\beta_R} \de_\gam t_{N} \prod_{i=1}^{N}\frac{ t_i^{m_i}}{1-t_i}\,,
 \end{split}
\end{equation}  
with normalisation constant
\begin{equation}
\begin{split}
 c_N=\int_{\beta_L}^{\gam\beta_R}\de_\gam t_1\int_{t_1}^{\gam\beta_R}\de_\gam t_2\cdots \int_{t_{N-1}}^{\gam\beta_R} \de_\gam t_{N} \prod_{i=1}^{N}\frac{1}{ (1-t_i)^{2}} \,.
 \end{split}
\end{equation} 
In the following section, we prove that \eqref{eq:stead} solves the stationarity condition \eqref{eq:Hmu}.

\section{Proof of the stationarity condition}\label{sec:steady} 
To proceed, we introduce  the  shorthand notation 
\begin{equation}\label{eq:intdef}
\int \mathcal{D}_\gam t^N:=\int_{\beta_L}^{\gam\beta_R} \mathrm{d}_\gam t_1 \int_{t_1}^{\gam\beta_R} \mathrm{d}_\gam t_2 \dots  \int_{t_{N-1}}^{\gam\beta_R} \mathrm{d}_\gam t_N \,,
\end{equation}
for the nested Jackson integral operator, cf.~\eqref{eq:stead}, such that the steady state is written as 
\begin{equation}\label{eq:stead2}
\begin{split}
 \mu (\vm) &=\frac{1}{c_N}\int \mathcal{D}_\gam t^N\prod_{i=1}^{N}\frac{ t_i^{m_i}}{1-t_i}\,.
 \end{split}
\end{equation}  
In the following, we provide the proof of stationarity \eqref{eq:Hmu}.

\subsection{Stationary measure as integral product state}\label{sec:ipa}
In order to act with the Hamiltonian \eqref{eq:ham},  we first introduce the space of states for each site of the process. To do so, it is convenient to define the vector
\begin{equation}\label{eq:Xvec}
    X(t)=\sum_{m=0}^\infty \frac{t^m}{1-t}|m\rangle\,,
\end{equation}  
such that the steady state can be written as
\begin{equation}\label{eq:ipa}
    |\mu\rangle =\frac{1}{c_N}\int \mathcal{D}_\gam t^N X(t_1)\otimes \ldots\otimes  X(t_N) \,.
\end{equation} 
To proceed with the verification of \eqref{eq:Hmu} we move the Hamiltonian inside the integral and act on the tensor product of $X$ vectors \eqref{eq:Xvec}.

We begin computing the action of the left boundary operator $B_L$ that solely acts on the leftmost vector $X(t_1)$ in \eqref{eq:ipa}. From the action \eqref{eq:bd1} and after shifting the sums, we find that
\begin{equation}\label{eq:BxL}
\begin{split} 
B_L X(t_1) =  \bar X(\beta_L,t_1)\,,
  \end{split}
\end{equation}
where we defined
\begin{equation}\label{eq:bX}
 \begin{split}
\bar X(s,t)=  \sum_{m =0}^\infty\; \frac{ t^{m}}{1-t}  \left(\sum _{k=1}^{m} \frac{\gamma^{k}}{\br{k}} -\sum_{k=1}^{m} \frac{1}{\br{k}}\left(\frac{s}{t}\right)^k+\sum_{k=1}^\infty \frac{s^k}{\br{k}}-\sum_{k=1}^\infty \frac{(\gam t)^k}{\br{k}}\right)
 |m \rangle \,.
\end{split}
\end{equation}

A similar procedure for the right boundary yields
\begin{equation}\label{eq:BXR}
\begin{split}
 B_R X(t_N)=\bar  Y(t_N,\beta_R)\,,
  \end{split}
\end{equation} 
with $\bar Y$ defined via
\begin{equation}\label{eq:bY}
 \begin{split}
\bar  Y(s,t)=  \sum_{m=0}^\infty \frac{ s^{m}}{1-s} \left(\sum _{k=1}^{m} \frac{1}{\br{k}} -\sum_{k=1}^{m} \frac{1}{\br{k}}\left(\frac{\gam t}{s}\right)^k+\sum_{k=1}^\infty \frac{(\gam t)^k}{\br{k}}-\sum_{k=1}^\infty \frac{s^k}{\br{k}}\right)
   |m \rangle\,.
\end{split}
\end{equation}

We now turn to the bulk action of the Hamiltonian. The Hamiltonian density only acts on two neighbouring sites. Using 
\begin{equation*}
 \begin{split}
  &  \sum_{m_i,m_{i+1}=0}^\infty t_i^{m_i} t_{i+1}^{m_{i+1}}\sum_{k=1}^{m_i}\frac{1}{\br{k}}|m_i-k\rangle\otimes |m_{i+1}+k\rangle 
  =
  \sum_{m_i,m_{i+1}=0}^\infty t_i^{m_i}  t_{i+1}^{m_{i+1}}\sum_{k=1}^{m_{i+1}}\frac{1}{\br{k}}\left(\frac{t_i}{t_{i+1}}\right)^k |m_i\rangle\otimes |m_{i+1}\rangle\,,
\end{split}
\end{equation*} 
and 
\begin{equation*}
 \begin{split}
  &  \sum_{m_i,m_{i+1}=0}^\infty  t_i^{m_i}  t_{i+1}^{m_{i+1}} \sum_{k=1}^{m_{i+1}}\frac{\gam^k}{\br{k}}|m_i+k\rangle\otimes |m_{i+1}-k\rangle
  =
  \sum_{m_i,m_{i+1}=0}^\infty  t_i^{m_i}  t_{i+1}^{m_{i+1}} \sum_{k=1}^{m_{i}}\frac{\gam^k}{\br{k}}\left(\frac{t_{i+1}}{t_{i}}\right)^k |m_i\rangle\otimes |m_{i+1}\rangle \,,
\end{split}
\end{equation*}
we obtain
\begin{equation}
 \begin{split} 
\mathcal{H}\left(X(t_i)\otimes X(t_{i+1}) \right)
&=  \sum_{m_i,m_{i+1}=0}^\infty  \frac{ t_i^{m_i}}{1-t_i}    \frac{ t_{i+1}^{m_{i+1}}}{1-t_{i+1}} \\ &\times   \left(\sum _{k=1}^{m_i} \frac{1}{\br{k}}-\sum_{k=1}^{m_{i}}\frac{\gam^k}{\br{k}}\left(\frac{t_{i+1}}{t_{i}}\right)^k+\sum _{k=1}^{m_{i+1}} \frac{\gamma^{k}}{\br{k}}-\sum_{k=1}^{m_{i+1}}\frac{1}{\br{k}}\left(\frac{t_i}{t_{i+1}}\right)^k\right)
|m_i\rangle\otimes |m_{i+1}\rangle\\
&=  X(t_i)\otimes \bar X(t_{i},t_{i+1}) +\bar Y(t_i,t_{i+1})\otimes X(t_{i+1})\,,
\end{split}
\end{equation}
with $\bar X$ and $\bar Y$ defined in \eqref{eq:bX} and \eqref{eq:bY}.

Thus, in order to show that the action of the Hamiltonian on the steady state \eqref{eq:stead2} vanishes, it remains to verify that 
\begin{equation}\label{eq:toshow}
 \sum_{i=1}^N\int \mathcal{D}_\gam t^N X(t_1)\otimes \ldots \otimes X(t_{i-1})\otimes\left[ \bar X(t_{i-1},t_i)+\bar Y(t_i,t_{i+1})\right]\otimes X(t_{i+1})\otimes \ldots \otimes X(t_{N})=0
\end{equation} 
where $t_0=\beta_L$ and $t_{N+1}=\beta_R$.

This expression resembles the structure of the matrix product ansatz \cite{Derrida:1992vu}. The term-by-term cancellation happens after integration, which is shown in the following section. 
 
 \subsection{Evaluation of the q-integrals}\label{sec:proof}

To proceed, it is convenient to introduce a polynomial vector space 
\begin{equation}
 \lambda_1^{m_1}\cdots \lambda_N^{m_N}\simeq |m_1\rangle\otimes\ldots\otimes|m_N\rangle\,,
\end{equation}  
with formal parameters $\lambda_i\in \mathbb{C}$.
Let us consider the projection by the covector  
\begin{equation}\label{eq:proje}
 \langle\lambda_i|=\sum_{m=0}^\infty\lambda_i^{m} \langle m|\,,
\end{equation} 
 at a given site $i$.
Then, its action on $\bar X(t_{i-1},t_i)+ \bar Y(t_i,t_{i+1})$ can be written as 
\begin{equation}
\begin{split}
  \langle\lambda_i|\bar X(t_{i-1},t_i)+  \langle\lambda_i|\bar Y(t_i,t_{i+1})  &=  \sum_{k=1}^\infty (1-\lambda_i^k)  \left( \frac{t_{i-1}^k+(\gam t_{i+1})^k}{\br{k}}  -  \frac{t_i^k+(\gam t_i)^k}{\br{k}} \right)
 \frac{1}{1-t_i}\frac{1}{1-\lambda_i t_i}\\
 &=  (1-\lambda_i)   \left(F_{\lambda_i}(t_{i-1})+F_{\lambda_i}(\gam t_{i+1}) -F_{\lambda_i}(t_i)-F_{\lambda_i}(\gamma t_i)   \right)
 f_{\lambda_i}(t_i)\,.
 \end{split}
\end{equation}   
Here, we defined  the function
\begin{equation} \label{eq:f}
f_\lambda(t)= \frac{1}{(1-t)(1-\lambda t)} \,,
\end{equation} 
along with its q-antiderivative  with respect to the argument $t$ 
\begin{equation}\label{eq:relati}
\begin{split}
F_\lambda(t)=  \frac{1}{1-\lambda}\sum_{k=1}^\infty\frac{t^k \left(1-\lambda ^{k}\right)}{ \br{k}}+C\,,
 \end{split}
\end{equation} 
cf.~\eqref{eq:qderivative} and \eqref{eq:qanti}, with the integration  constant $C$.

It then follows that after multiplication \eqref{eq:toshow} with $\langle\lambda_1|\otimes\ldots\otimes \langle\lambda_N|$ we get
\begin{equation}\label{eq:master_transformed2}
\begin{split} 
\sum_{j=1}^N (1-\lambda_j)
\int \mathcal{D}_\gam t^N   \left[\prod_{\ell=1}^N f_{\lambda_\ell}(t_\ell) \right] &\left[F_{\lambda_j}(t_{j-1})+F_{\lambda_j}(\gam t_{j+1})-F_{\lambda_j}(t_{j})-F_{\lambda_j}(\gam t_{j})\right] =0\,.
\end{split}
\end{equation} 
For $\gamma\to 1$, we recover the rational case as considered in \cite{2023arXiv230702793C}. A less intuitive but equivalent way to derive \eqref{eq:master_transformed2} is to introduce the $\lambda$ parameters right from the beginning and collect terms containing only $\lambda_i$ corresponding to a given site $i$. In this way, one arrives at the same result, but the analogy with the matrix product ansatz is lost. For completeness, we present it in Appendix~\ref{app:genf}. 

Let us now evaluate the Jackson integrals in \eqref{eq:master_transformed2}.
We first consider the  Jackson integral in $t_N$ over the last term in the sum of the integrand \eqref{eq:master_transformed2}, i.e. 
\begin{equation}
(1-\lambda_N)\int_{t_{N-1}}^{\gam t_{N+1}}\mathrm{d}_\gam t_N f_{\lambda_N}(t_N) \left[F_{\lambda_N}(t_{N-1})+F_{\lambda_N}(\gam t_{N+1})-F_{\lambda_N}(t_{N})-F_{\lambda_N}(\gam t_{N})\right]\,.
\end{equation}
Using the q-analog of integration by parts, see~\eqref{eq:IPf}, we observe that this Jackson integral above vanishes
\begin{equation}
 \left(F_{\lambda_N}(t_{N-1})+F_{\lambda_N}(\gam t_{N+1})\right) \left(F_{\lambda_N}(\gam t_{N+1})-F_{\lambda_N}(t_{N-1})\right)- \left(   F_{\lambda_N}^2(\gam t_{N+1})  -   F_{\lambda_N}^2(t_{N-1})      \right)=0\,.
\end{equation}
We remark that for $N=1$, the vanishing of the integral is equivalent to the proof given in Appendix~\ref{app:l1} expressed through infinite sums.

We now focus on the Jackson integral over the $k^{\text{th}}$ term with $k<N$ in the sum of the integrand in \eqref{eq:master_transformed2}. Similar to \eqref{eq:intdef}, we introduce the notation
\begin{equation}
\int  \mathcal{D} _\gam t^{k-1}=\int_{\beta_L}^{\gam t_{N+1}} \mathrm{d}_\gam t_1 \int_{t_1}^{\gam t_{N+1}} \mathrm{d}_\gam t_2 \dots  \int_{t_{k-2}}^{\gam t_{N+1}} \mathrm{d}_\gam t_{k-1}    \,,
\end{equation}
for the nested Jackson q-integral operator and also the function
\begin{equation}
L_{k+1}(t_{k+1})=\int_{t_{k+1}}^{\gam t_{N+1}} \mathrm{d}_\gam t_{k+2} \int_{t_{k+2}}^{\gam t_{N+1}} \mathrm{d}_\gam t_{k+3} \dots  \int_{t_{N-1}}^{\gam t_{N+1}} \mathrm{d}_\gam t_{N} \prod_{j=k+2}^N f_{\lambda_j}(t_j)\,.
\end{equation}
With these, the $k^{\text{th}}$ term in the sum of \eqref{eq:master_transformed2} can then be written as
\begin{equation}\label{eq:kth_term_integral}
\begin{split}
(1-\lambda_k) \int   {\mathcal{D}}_\gam t^{k-1}\left[\prod_{j=1}^{k-1} f_{\lambda_j}(t_j) \right] & \int_{t_{k-1}}^{\gam t_{N+1}} \mathrm{d}_\gam t_k   \int_{t_{k}}^{\gam t_{N+1}} \mathrm{d}_\gam t_{k+1}    
 f_{\lambda_k}(t_k)  f_{\lambda_{k+1}}(t_{k+1}) \\
 &\times\left[F_{\lambda_k}(t_{k-1})+F_{\lambda_k}(\gam t_{k+1})-F_{\lambda_k}(t_{k})-F_{\lambda_k}(\gam t_{k})\right] L_{k+1}(t_{k+1})\,.
 \end{split}
\end{equation}
Let us focus on the integration over the $t_k$ and $t_{k+1}$ variables. For ordinary integration when $\gamma\to 1$, one can change the integration limits according to 
\begin{equation}
\int_{t_{k-1}}^{t_{N+1}}\mathrm{d}t_{k} \int_{t_{k}}^{t_{N+1}}\mathrm{d}t_{k+1}\, g(t_{k},t_{k+1})=\int_{t_{k-1}}^{t_{N+1}}\mathrm{d}t_{k+1} \int_{t_{k-1}}^{t_{k+1}}\mathrm{d}t_{k}\, g(t_k,t_{k+1})\,,
\end{equation}
for a generic function $g$. However, for the q-deformed Jackson integral, we have a correction term, i.e.
\begin{equation}\label{eq:int_rearrange}
\begin{split}
\int_{t_{k-1}}^{\gam t_{N+1}}\mathrm{d}_\gam t_{k} \int_{t_{k}}^{\gam t_{N+1}}\mathrm{d}_\gam t_{k+1}\, g(t_{k},t_{k+1})&=\int_{t_{k-1}}^{\gam t_{N+1}}\mathrm{d}_\gam t_{k+1} \int_{t_{k-1}}^{ t_{k+1}}\mathrm{d}_\gam t_{k}\, g(t_{k},t_{k+1})\\
&\qquad -(1-\gam) \int_{t_{k-1}}^{\gam t_{N+1}}\mathrm{d}_\gam t_{k+1}\,   g(t_{k+1},t_{k+1})\, t_{k+1}\,.
\end{split}
\end{equation}  
This relation can be explicitly shown by writing the q-integrals as infinite sums like in the definition of the Jackson integral \eqref{eq:jack1} and \eqref{eq:jack2}. The extra term vanishes in the limit $\gam \to 1$. 

By performing the exchange of the integration limits in \eqref{eq:kth_term_integral} and evaluating the $t_k$ integral by using \eqref{eq:qanti} and \eqref{eq:IPf}, we find that 
\begin{equation}\label{eq:lasteq} 
\begin{split}
&\int_{t_{k-1}}^{\gam t_{N+1}} \mathrm{d}_\gam t_k   \int_{t_{k}}^{\gam t_{N+1}} \mathrm{d}_\gam t_{k+1}    
 f_{\lambda_k}(t_k)  f_{\lambda_{k+1}}(t_{k+1}) \left[F_{\lambda_k}(t_{k-1})+F_{\lambda_k}(\gam t_{k+1})-F_{\lambda_k}(t_{k})-F_{\lambda_k}(\gam t_{k})\right] L_{k+1}(t_{k+1})\\
 &=\int_{t_{k-1}}^{\gam t_{N+1}} \mathrm{d}_\gam t_{k+1}  f_{\lambda_{k+1}}(t_{k+1}) \left[F_{\lambda_k}(t_{k-1})-F_{\lambda_k}(t_{k+1})\right] L_{k+1}(t_{k+1}) \\
 &\quad  \times \Bigg\{  (F_{\lambda_k}(t_{k+1})- F_{\lambda_k}(\gam t_{k+1})  - (1-\gam)  t_{k+1}  f_{\lambda_{k}}(t_{k+1})   \Bigg\}\,.
 \end{split}
\end{equation}
The expression between the curly brackets vanishes due to the definition of the q-derivative \eqref{eq:qderivative}; hence,  so does all the $j=1, \dots, N-1$ terms in the stationary equation \eqref{eq:master_transformed2}. This ends the proof that \eqref{eq:stead} is indeed the stationary solution of the MADM defined in \eqref{eq:gen}.

\section{Equilibrium stationary measure}\label{sec:equi}

In this section, we evaluate the stationary measure for the equilibrium case, i.e. $\beta=\beta_R=\beta_L$. More precisely, we show that  for this particular choice of boundary parameters, it becomes the product measure of geometric distributions
\begin{equation}\label{eq:reseq}
    \mu^{\text{eq}}(\vm)= \prod_{i=1}^N\beta^{m_i}(1-\beta) \,.
\end{equation}
For this purpose, we first examine the behaviour of the non-normalised stationary measure defined via
\begin{equation}
    \tilde \mu_N (\vm)=c_N \mu_N (\vm)\,,
\end{equation}
where we introduced the $N$ subscript of the steady state to emphasize the length $N$ of the system on which the measure is defined. 
As direct consequence of the definition   \eqref{eq:leng1}, we obtain 
\begin{equation}
    \tilde \mu^{\text{eq}}_1(m_1)=(\gamma-1)\frac{\beta^{m_1+1}}{1-\beta}\,,
\end{equation}
for $N=1$. 
For general $N$, we have the product structure
\begin{equation}\label{eq:mutilde_eq}
   \tilde\mu_N^{\text{eq}}(\vm) =(\gam-1)^N\prod_{i=1}^N\frac{\beta^{m_i+1}}{1-\beta}\,.
\end{equation}
This is proved by the method of induction below. 

In order to prove \eqref{eq:mutilde_eq}, let us introduce the auxiliary function
\begin{equation}
    \phi_m(\beta)=\sum_{k=m+1}^\infty \frac{\beta^k}{\br{k}}\,,
\end{equation}
to express the Jackson integral as  
\begin{equation}
    \int_w^{\gam\beta}\de_\gam t\frac{t^m}{1-t} =  \phi_{m}(\gam\beta)-\phi_{m}(w) \,.
\end{equation}
With this notation, the definition of the steady state \eqref{eq:stead} implies the following recursion relation for $N>1$:
\begin{equation}\label{eq:recur}
\begin{split}
   &  \tilde \mu_N (m_1,\ldots,m_N)=\tilde \mu_{N-1} (m_1,\ldots,m_{N-1})\phi_{m_N}(\gam\beta_R)-\sum_{k=m_N+1}^\infty\frac{1}{\br{k}}\tilde \mu_{N-1}(m_1,\ldots,m_{N-1}+k)\,. 
    \end{split}
\end{equation}
Let us assume the form \eqref{eq:mutilde_eq} is valid for $\tilde\mu_{N-1}^{\text{eq}}(m_1,\ldots,m_{N-1})$. Plugging it into \eqref{eq:recur} leads to
\begin{equation}
\begin{split}
  \tilde \mu_N^{\text{eq}} (m_1,\ldots,m_N)&= (\gam-1)^{N-1}\prod_{i=1}^{N-1}\frac{\beta^{m_i+1}}{1-\beta}\left(\phi_{m_N}(\gam\beta)-\sum_{k=m_N+1}^\infty\frac{1}{\br{k}}\beta^{k}\right)\\
  &=(\gam-1)^{N-1}\prod_{i=1}^{N-1}\frac{\beta^{m_i+1}}{1-\beta}\left(\phi_{m_N}(\gam\beta)-\phi_{m_N}(\beta)\right) \\
  &=(\gam-1)^{N-1}\prod_{i=1}^{N-1}\frac{\beta^{m_i+1}}{1-\beta}\left((\gam-1)\frac{\beta^{m_N+1}}{1-\beta}\right)=(\gam-1)^{N}\prod_{i=1}^{N}\frac{\beta^{m_i+1}}{1-\beta}\,;
    \end{split}
\end{equation}
that is exactly of the form \eqref{eq:mutilde_eq}. Since we already showed the validity of the base case where $N=1$, the induction is complete and \eqref{eq:mutilde_eq} is valid for general $N$.

The normalisation of the stationary measure is then computed via
\begin{equation}
    c_N^{\text{eq}}=\sum_{m_1,m_2\dots,m_N=0}^{\infty}\tilde \mu_N^{\text{eq}} (m_1,\ldots,m_N)= (\gam-1)^N\frac{\beta^{N}}{\left(1-\beta\right)^{2N}}\,,
\end{equation}
and hence, we obtain
\begin{equation}
    \mu_N^{\text{eq}}(\vm)=\frac{1}{c_N^{\text{eq}}} \tilde\mu_N^{\text{eq}}(\vm)=\prod_{i=1}^N\beta^{m_i}(1-\beta) \,,
\end{equation}
cf.~\eqref{eq:reseq}.

\section{Conclusion} \label{sec:con}
In this note, we gave an exact expression for the steady state of the boundary-driven multiparticle asymmetric diffusion model (MADM) and computed the product measure at equilibrium where $\beta=\beta_R=\beta_L$. 
The obtained results rely to some extent on inspiration taken from the rational limiting case. The formula for the steady state presented in \eqref{eq:stead} beautifully encodes the rational case where the Jackson q-integral will turn into the ordinary integral.

Although leading to the desired result, the derivation presented here is somewhat dissatisfying. The reader may have noticed that we had to make two ans\"atze in order to derive the stationary measure. The first concerns the stationary measure at length $N=1$ in Section~\ref{sec:ansatz}, and the second is the nested structure for arbitrary $N$ in Section~\ref{sec:steady}.  To gain further insights, it would be interesting to reformulate our results in terms of a suitable, quasi-local representation of the matrix product ansatz \cite{Derrida:1992vu} that arises from the underlying Zamolodchikov algebra, see \cite{doi:10.1143/JPSJ.66.2618,Crampe:2014aoa}. The existence of such formulation is further motivated by the observation that in the proof presented in Section~\ref{sec:proof}, only two neighbouring integrals play a nontrivial role.  We shall come back to this point in a follow-up work.
 
 \subsection*{Acknowledgements}
We thank Gioia Carinci, Cristian Giardin\`a, Rob Klabbers and Frank Redig for useful discussions.
We acknowledge support from the INFN grant "Gauge and String Theory (GAST)",  the “INdAM–GNFM Project”, CUP-E53C22001930001, the FAR UNIMORE project CUP-E93C23002040005 and the PRIN project "2022ABPBEY", CUP-E53D23002220006. IMSZ  received support from Nordita that is supported in part by NordForsk.  We thank the anonymous referees for their comments on the manuscript.

\appendix

  \section{Basic q-calculus}\label{app:qcalc}
The q-derivative 
of a function $G(x)$ is defined via
\begin{equation} \label{eq:qderivative}
D_\gam G(x)=g(x)=\frac{G(\gam x)-G(x)}{\gam x-x}\,.
\end{equation} 
 The Jackson integral of  $g$ is expressed in terms of its q-antiderivative $G$ via 
\begin{equation}\label{eq:qanti}
G(b)-G(a)=\int_a^b\de_\gam t \,g(t)\,.
\end{equation} 
The q-analog of the integration by parts can be found in \cite{kac2002quantum}. For two functions $G$ and $H$ we have
\begin{equation}
 \int_{a}^b H(t)(D_\gam G(t))\de_\gam t=H(b)G(b)-H(a)G(a)-\int_{a}^b G(\gam t)(D_\gam H(t)) \de_\gam t\,.
\end{equation}
For the special case where $G=H$, we get 
\begin{equation}\label{eq:IPf}
 \int_{a}^b( G(t)+G(\gam t))(D_\gam G(t))\de_\gam t=G(b)^2-G(a)^2\,.
\end{equation}

\section{Proof of the $N=1$ sum formula of the stationary state}\label{app:l1} 

Our strategy to prove the stationary state formula for $N=1$ is to plug in the ansatz \eqref{eq:leng1}
\begin{equation}
 \mu_\gam(m)=c\sum_{k=m+1}^\infty \frac{(\gam\beta_R)^k-\beta_L^k}{\br{k}}\,, 
\end{equation} 
into the stationary equation \eqref{eq:stateq}
\begin{equation}
\begin{split}
 \left(\sum _{k=1}^{m} \frac{1+\gam^{k}}{\br{k}}+\sum_{k=1}^\infty\frac{(\gam\beta_R)^k+\beta_L^k}{\br{k}}\right)\mu(m)&=
\sum_{k=1}^\infty 
\frac{1+\gam ^{k} }{\br{k}}\mu(m+k) +\sum_{k=1}^m\frac{(\gam\beta_R)^k+\beta_L^k}{\br{k}}\mu(m-k) \,,
\end{split}
\end{equation}  
and show that coefficient of every $\beta_L^{p}\beta_{R}^{q}$ term vanish. 

First, let us focus on the terms that have both $\beta_R$ and $\beta_L$ dependence. Moving all the terms onto one side of the stationary equation, such terms are
\begin{equation}
\sum_{k=1}^\infty\sum_{l=m+1}^\infty \frac{(\gam\beta_R)^k \beta_L^l-(\gam\beta_R)^l \beta_L^k}{\br{k}\br{l}} -\sum_{l=1}^m \sum_{k=m-l+1}^\infty \frac{(\gam\beta_R)^l \beta_L^k-(\gam\beta_R)^k \beta_L^l}{\br{k}\br{l}}\,.
\end{equation}
With elementary manipulation, we can show
\begin{equation}
    \begin{split}
    & \sum_{k=1}^\infty\sum_{l=1}^\infty \frac{(\gam\beta_R)^k \beta_L^l-(\gam\beta_R)^l \beta_L^k}{\br{k}\br{l}}-\sum_{k=1}^\infty\sum_{l=1}^m \frac{(\gam\beta_R)^k \beta_L^l-(\gam\beta_R)^l \beta_L^k}{\br{k}\br{l}} -\sum_{l=1}^m \sum_{k=m-l+1}^\infty \frac{(\gam\beta_R)^l \beta_L^k-(\gam\beta_R)^k \beta_L^l}{\br{k}\br{l}} \\
    =&\sum_{k=1}^\infty\sum_{l=1}^\infty \frac{(\gam\beta_R)^k \beta_L^l-(\gam\beta_R)^l \beta_L^k}{\br{k}\br{l}} -\sum_{l=1}^m \sum_{k=1}^{m-l}\frac{(\gam\beta_R)^k \beta_L^l-(\gam\beta_R)^l \beta_L^k}{\br{k}\br{l}}\,,
\end{split}
\end{equation}
where both the double sums vanish due to symmetry reasons.

With the mixed terms vanishing, the rest of the stationary equation breaks into two equations containing only $\gam\beta_R$ or only $\beta_L$ dependence. They are both proportional to
\begin{equation}
\sum_{l=m+1}^\infty \frac{\beta^l}{\br{l}} \sum _{k=1}^{m} \frac{1+\gam^{k}}{\br{k}}  +\sum_{k=1}^\infty \sum_{l=m+1}^\infty \frac{\beta^{k+l}}{\br{k}\br{l}} -\sum_{k=1}^\infty 
\frac{1+\gam ^{k} }{\br{k}}\sum_{l=m+k+1}^\infty \frac{\beta^l}{\br{l}} -\sum_{k=1}^m\sum_{l=m-k+1}^\infty\frac{\beta^{k+l}}{\br{k}\br{l}} \,,   
\end{equation}
where $\beta=(\gam\beta_R)$ or $\beta=\beta_L$. Reshuffling the summation indices leads to the form
\begin{equation}
\sum_{l=m+1}^\infty \frac{\beta^l}{\br{l}} \sum _{k=1}^{m} \frac{1+\gam^{k}}{\br{k}}
+\sum_{l=m+2}^\infty\sum_{k=1}^{l-m-1}  \frac{\beta^{l}}{\br{k}\br{l-k}} -\sum_{l=m+2}^\infty\sum_{k=1}^{l-m-1} 
\frac{1+\gam ^{k} }{\br{k}} \frac{\beta^l}{\br{l}} -\sum_{l=m+1}^\infty\sum_{k=1}^m\frac{\beta^{l}}{\br{k}\br{l-k}} \,.  
\end{equation}
To continue, we need to distinguish different cases according to the power of $\beta^p$. 

In case $p=m+1$, the coefficient is 
\begin{equation}
   \frac{1}{\br{m+1}} \sum _{k=1}^{m} \left( \frac{1+\gam^{k}}{\br{k}}-\frac{\br{m+1}}{\br{k}\br{m+1-k}}\right)=\frac{1}{\br{m+1}} \sum _{k=1}^{m} \left( \frac{1}{\br{k}}-\frac{1}{\br{m+1-k}}\right)\,,
\end{equation}
where we used the definition of the q-number \eqref{eq:qnum}. The sum vanishes due to symmetry reasons.

In case $m+2\leq p \leq 2m $, the coefficient of $\beta^{p}$ is
\begin{equation}
   \frac{1}{\br{p}} \sum_{k=p-m}^{m} \left( \frac{1+\gam^{k}}{\br{k}}-\frac{\br{p}}{\br{k}\br{p-k}}\right)=\frac{1}{\br{p}} \sum_{k=p-m}^{m} \left( \frac{1}{\br{k}}-\frac{1}{\br{p-k}}\right)\,,
\end{equation}
that also vanishes due to symmetry reasons.

For $p=2m+1$, the different terms in the coefficients cancel out each other automatically. 

For $2m+2\leq p$, we have 
\begin{equation}
   \frac{1}{\br{p}} \sum_{k=m}^{p-m} \left( \frac{1+\gam^{k}}{\br{k}}-\frac{\br{p}}{\br{k}\br{p-k}}\right)=\frac{1}{\br{p}} \sum_{k=m}^{p-m} \left( \frac{1}{\br{k}}-\frac{1}{\br{p-k}}\right)\,,
\end{equation}
that is zero, similar to the previous terms. 

Since we showed that every coefficient of $\beta_L^{p}\beta_{R}^{q}$ vanishes, we proved that our ansatz is the stationary state for length $N=1$.

 \section{Generating function method} \label{app:genf}

Let us consider the projection 
\begin{equation}\label{eq:genfun}
\sum_{m_1,\ldots,m_N=0}^\infty  \lambda_1^{m_1}\cdots\lambda_N^{m_N} \langle\vm|H|\mu \rangle=0\,,
\end{equation}
cf.~\eqref{eq:proje}, 
and recall the definition of $f_\lambda$ in \eqref{eq:f}. The following relations hold for arbitrary coefficients $a_k$:
\begin{equation}\label{eq:relss}
\begin{split}
\sum_{m_1,\ldots,m_N=0}^\infty \lambda_1^{m_1}\cdots\lambda_N^{m_N} \mu_\gam(m)&= \int \mathcal{D}_\gam t ^N\prod_{j=1}^N f_{\lambda_j}(t_j)\,,\\
\sum_{m_1,\ldots,m_N=0}^\infty \lambda_1^{m_1}\cdots\lambda_N^{m_N} \sum_{k=1}^{m_\ell}a_k\,\mu_\gam(m)&= \int \mathcal{D}_\gam t^N      \sum_{k=1}^\infty a_k\, \lambda_\ell^k\, t_\ell^k\prod_{j=1}^N f_{\lambda_j}(t_j)   \,,\\
\sum_{m_1,\ldots,m_N=0}^\infty \lambda_1^{m_1}\cdots\lambda_N^{m_N} \sum_{k=1}^{m_\ell}a_k\,\mu_\gam(m-k\delta_\ell)&= \int \mathcal{D}_\gam t^N    \sum_{k=1}^\infty a_k\, \lambda_\ell^k\prod_{j=1}^N f_{\lambda_j}(t_j)  \,,\\
\sum_{m_1,\ldots,m_N=0}^\infty \lambda_1^{m_1}\cdots\lambda_N^{m_N}\sum_{k=1}^{\infty}a_k\,\mu_\gam(m+k\delta_\ell)&= \int \mathcal{D}_\gam t^N \sum_{k=1}^\infty a_k \, t_\ell^k  \prod_{j=1}^N f_{\lambda_j}(t_j)   \,,\\
\sum_{m_1,\ldots,m_N=0}^\infty \lambda_1^{m_1}\cdots\lambda_N^{m_N} \sum_{k=1}^{m_\ell}a_k\,\mu_\gam(m-k\delta_\ell+k\delta_{\ell+1})&= \int \mathcal{D}_\gam t^N     \sum_{k=1}^\infty a_k \,\lambda_\ell^k \,t_{\ell+1}^k \prod_{j=1}^N f_{\lambda_j}(t_j)   \,,\\
\sum_{m_1,\ldots,m_N=0}^\infty \lambda_1^{m_1}\cdots\lambda_N^{m_N} \sum_{k=1}^{m_{\ell+1}}a_k\,\mu_\gam(m+k\delta_\ell-k\delta_{\ell+1})&= \int \mathcal{D}_\gam t^N     \sum_{k=1}^\infty a_k \,\lambda_{\ell+1}^k \,t_{\ell}^k  \prod_{j=1}^N f_{\lambda_j}(t_j)   \,,
\end{split}
\end{equation} 
for $1\leq \ell\leq N$.
These relations allow us to rewrite \eqref{eq:genfun} as 
\begin{equation}
\begin{split}
\int \mathcal{D}_\gam t^N  \left[\prod_{\ell=1}^N f_{\lambda_\ell}(t_\ell) \right]  \sum_{k=1}^\infty\frac{1}{\br{k}} & \Bigg[ \beta_L^k (1-\lambda_1^k)+ (\gam\beta_R)^k (1-\lambda_N^k) -\left(1-\lambda_N^k\right)t_N^k-\gam^k\left(1-\lambda_1^k\right)t_1^k
 \\
&+\sum_{j=1}^{N-1}\left(\lambda_j^k-\lambda_{j+1}^k\right)t_j^k+\sum_{j=2}^{N}\gam^k\left(\lambda_j^k-\lambda_{j-1}^k\right)t_{j}^k \Bigg]=0\,.
\end{split}
\end{equation} 
Further inserting a zero, i.e. $0=1-1$, into the brackets of the second line above and collecting the terms $(1-\lambda_i^k)$, we get
\begin{equation}
\begin{split}
\int \mathcal{D}_\gam t^N  \left[\prod_{\ell=1}^N f_{\lambda_\ell}(t_\ell) \right] \sum_{k=1}^\infty\frac{1}{\br{k}} & \Bigg[ (1-\lambda_1^k)(\beta_L^k-t_1^k-\gam^kt_1^k+\gam^kt_2^k) 
 \\
&+\sum_{j=2}^{N-1}\left(1-\lambda_j^k\right) (t_{j-1}^k-t_j^k-\gam^kt_j^k+\gam^kt_{j+1}^k) \\
&+ \left(1-\lambda_N^k\right)((\gam\beta_R)^k -t_N^k-\gam^kt_N^k+t_{N-1}^k)\Bigg]=0\,.
\end{split}
\end{equation}  
The integrand can then be expressed in terms of the q-antiderivative \eqref{eq:relati} such that we find exactly \eqref{eq:master_transformed2}. We remark that for $\gamma\to 1$, i.e. the rational case, this path reduces to the calculation presented in \cite{2023arXiv230702793C}.

\bibliography{refs.bib}
\bibliographystyle{utphys2.bst}
\end{document}